\documentclass[aps,prd,showpacs,preprintnumbers,nofootinbib]{revtex4}

\usepackage[dvipdfmx]{graphicx}
\usepackage{amsmath,amssymb}
\usepackage{bm}
\usepackage{color}

\newcommand{\nnmb}{\nonumber\\}

\begin{document}

\preprint{YITP-11-88}
\preprint{ICRR-Report-596-2011-13}

\author{Kazuyuki Sugimura$^1$}
\email{sugimura@yukawa.kyoto-u.ac.jp}
\author{Daisuke Yamauchi$^{1, 2}$}
\email{yamauchi@icrr.u-tokyo.ac.jp}
\author{Misao Sasaki$^1$}
\email{misao@yukawa.kyoto-u.ac.jp}
\affiliation{%
$^1$Yukawa Institute for Theoretical Physics, Kyoto University, Kyoto, Japan\\
$^2$Institute for Cosmic Ray Research, University of Tokyo, Chiba, Japan
}%

\title{Multi-field open inflation model\\
  and\\
  multi-field dynamics in tunneling
  }% Force line breaks with \\

\date{\today}% It is always \today, today,
  % but any date may be explicitly specified

\begin{abstract}
We consider a multi-field open inflation model, in which
one of the fields dominates quantum tunneling from a false vacuum
while the other field governs slow-roll inflation within 
the bubble nucleated from false vacuum decay. We call the former the
tunneling field and the latter the inflaton field.
In the limit of a negligible interaction between the two fields,
the false vacuum decay is described by a Coleman-De Luccia instanton.
Here we take into account the coupling between the two fields
and construct explicitly a multi-field instanton for 
a simple quartic potential model. We also solve the evolution of the
scalar fields within the bubble. We find our model 
realizes open inflation successfully. This is the first concrete,
viable model of open inflation realized with a simple potential.
We then study the effect of the multi-field dynamics on
the false vacuum decay, specifically on the tunneling rate.
We find the tunneling rate increases in general provided that
the multi-field effect can be treated perturbatively.

\end{abstract}
\pacs{98.80.Cq, 04.62.+v}% PACS, the Physics and Astronomy
                             % Classification Scheme.

\maketitle
  %%%%%%%%%%%%%%%%%%%%%%%%%%%%%%%%%%%%%%%%%%%%%%%%%%%%%%%%%%%%%%%%%%%% 
 \section{Introduction}
 \label{sec:introduction}
 %%%%%%%%%%%%%%%%%%%%%%%%%%%%%%%%%%%%%%%%%%%%%%%%%%%%%%%%%%%%%%%%%%%% 
 %%%%%%%%%%%%%%%%%%%%%%%%%%%%%%%%%%%%%%%%%%%%%%%%%%%%%%%%%%%%%%%%%%%%

Vacuum decay is one of the most intriguing phenomena in field theory.
It occurs via the nucleation of a vacuum bubble by quantum tunneling 
from a metastable vacuum~\cite{Coleman:1977py}.
When gravity is included, it was first studied
by Coleman and De Luccia~\cite{Coleman:1980aw} who argued that
the tunneling is mediated by an $O(4)$-symmetric solution
of Euclidean Einstein-scalar field equations, 
called a Coleman-De Luccia (CDL) instanton.
It may be noted that the CDL instanton is obtained for
a single real scalar field with potential having one local minimum 
and one global minimum.
We also note that, in the presence of gravity, there is
even a possibility of tunneling up from a more stable vacuum of lower
energy to a less stable vacuum of higher energy.
Though there is no direct evidence that our part of the universe is
inside one of those nucleated bubbles, there are theoretical reasons 
to believe that it may be actually the case.
In fact, in the context of the string 
landscape~\cite{Kachru:2003aw,Susskind:2003kw,Freivogel:2004rd}, 
it is argued that our part of the universe appeared as a result of
quantum tunneling after trapped in one of many metastable vacua of
string theory. 

Reflecting the $O(4)$-symmetry of a CDL instanton, the region inside the
bubble has $O(3,1)$-symmetry, and hence becomes an open 
Friedmann-Lema\^itre-Robertson-Walker (FLRW) universe.
Subsequently slow-roll inflation may occur inside the bubble.
If the duration of inflation is not too long, that is,
if the number of $e$-folds of inflation is about 50 or 60,
the universe becomes almost flat but not extremely flat.
The spatial curvature today may not be negligible.
It was indeed suggested that string landscape would prefer
a non-negligible spatial curvature today~\cite{Freivogel:2005vv}.
This scenario is often called one-bubble open 
inflation~\cite{Bucher:1994gb,Sasaki:1994yt}.

Since one-bubble open inflation can be regarded as an outcome of string
landscape, studies of open 
inflation~\cite{Linde:1995xm,Yamamoto:1995sw,Yamamoto:1996qq,Sasaki:1996qn,
GarciaBellido:1997te}
 are in a sense studies of string landscape,
and testing open inflation against observations
implies testing string landscape.
Recently, Yamauchi et al.~\cite{Yamauchi:2011qq}
have studied single-field open inflation in this context
and argued that the current observational data already constrain
the shape of the potential substantially provided that
the curvature parameter today is $\Omega_{K,0}=10^{-2}\sim10^{-3}$.

In most of previous studies on open inflation, it was assumed that
a single scalar field governs both the quantum tunneling and 
the subsequent evolution inside the bubble. Then a successful model
can be constructed only for a very artificial form of the potential,
because the condition for the realization of tunneling through a CDL instanton
and that for the subsequent slow-roll inflation are not easily satisfied
at the same time~\cite{Linde:1998iw,Yamauchi:2011qq}.
To see this difficulty, let us consider a simple single-field model
with a quartic potential,
\begin{align}
V(\sigma )
=\frac{m^2}{2}\sigma^{2}-\frac{\delta}{3}\sigma^{3}+\frac{\lambda}{4}\sigma^{4},
\label{Vsigma}
\end{align}
where $m^2$, $\delta$, $\lambda>0$. This potential
has a metastable minimum at $\sigma>0$ and stable minimum at $\sigma=0$.
We assume slow-roll inflation to occur at $\sigma >m_{pl}$, 
where $m_{pl}$ denotes the reduced Planck mass ($m_{pl}\equiv(8\pi G)^{-1/2}$),
that is, we assume chaotic inflation to take place inside the bubble.
Then it is easy to see that the condition for the slow-roll inflation 
require a broad potential barrier. But this implies that
the tunneling is not mediated by a CDL instanton but 
by a Hawking-Moss (HM)
instanton~\cite{Hawking:1981fz,Jensen:1983ac}.
Then the condition that inflation should last only 50 or 60 $e$-folds
renders the amplitude of the curvature perturbation too large.

In contrast, in the multi-field case, one can introduce
two or more different fields and let each field play each
specific role. In the two-field case, we can introduce a tunneling
field that governs the tunneling dynamics and an inflaton field
that realizes slow-roll inflation after tunneling.
In fact, a multi-field situation seems well motivated from
the viewpoint of string landscape since one naturally expects
the presence of a large number of scalar fields there.

In this paper, we focus on a two-field system where tunneling occurs 
in a similar way as the CDL instanton case.
Extending the CDL method, we solve the Euclidean Einstein-scalar
equations with appropriate boundary conditions.
As in the usual CDL case, the resulting instanton
solution tells us about the dynamics of vacuum decay,
the decay rate, and the state of the universe right after the tunneling.

Specifically, we consider a simple two-field theory with
a quartic potential and numerically solve the field equations
to obtain a multi-field CDL instanton. The coupling between the
two fields is assumed to be small but non-negligible. 
Namely, the false vacuum decay is dominated by the tunneling
field but the effect of the inflaton field on the tunneling path
is non-negligible.
With this instanton in hand, we solve the subsequent
evolution of the universe inside the bubble.
We find we can obtain a successful model of open inflation
for reasonable values of the model parameters without fine-tuning
except for a mild tuning of one of the parameters to satisfy the condition
that the number of $e$-folds of inflation inside the bubble be about 50 or 60.
We also find that there is non-trivial evolution of the tunneling 
field though its contribution to the cosmic expansion is always sub-dominant.

Then in order to understand the nature of the multi-field tunneling, 
we study how the multi-field dynamics affects the tunneling rate.
For this purpose, we consider a wider range of the model parameters 
which are not necessarily realistic. We find the tunneling rate
increases in general as the effect of the coupling becomes stronger.

Recently, Aguirre, Johnson and Larfors studied
 tunneling in multi-field systems in the context of the 
string compactification and the string 
landscape~\cite{Johnson:2008vn,Aguirre:2009tp}.
They claim that tunneling by an O(4)-symmetric instanton can be totally prohibited
when a dilatonic coupling between the tunneling field and the other fields
is large. Their conclusion seems contradictory to ours at first glace.
However, it is not so because they considered such a strong
dilatonic coupling that it modifies the instanton solution
completely, while we consider the case where the multi-field effect
can be treated perturbatively.

This paper is organized as follows.
In Sec.~\ref{sec:formulation} we formulate a multi-field instanton 
method with gravity, by straightforwardly extending the CDL instanton method.
In Sec.~\ref{sec:model} we construct a concrete two-field open inflation 
model with a simple quartic potential by solving for
a multi-field CDL instanton numerically 
and evolving the scalar fields inside the bubble after tunneling.
In Sec.~\ref{sec:probabilty} we study how the multi-field dynamics 
affects the tunneling rate.
Section~\ref{sec:coclusion} is devoted to conclusion and discussion.

%%%%%%%%%%%%%%%%%%%%%%%%%%%%%%%%%%%%%%%%%%%%%%%%%%%%%%%%%%%%%%%%%%%% 
%%%%%%%%%%%%%%%%%%%%%%%%%%%%%%%%%%%%%%%%%%%%%%%%%%%%%%%%%%%%%%%%%%%% 
\section{formulation}
\label{sec:formulation}
%%%%%%%%%%%%%%%%%%%%%%%%%%%%%%%%%%%%%%%%%%%%%%%%%%%%%%%%%%%%%%%%%%%% 
%%%%%%%%%%%%%%%%%%%%%%%%%%%%%%%%%%%%%%%%%%%%%%%%%%%%%%%%%%%%%%%%%%%%
Let us first formulate a method to describe multi-field tunneling,
extending the CDL instanton method for single-field tunneling 
with gravity~\cite{Coleman:1980aw}.
To be specific, we consider a system with two scalar fields, 
$\sigma$ and $\phi$, with potential $V(\sigma,\phi)$.
We assume this potential has one global minimum $(\sigma_T,\phi_T)$
corresponding to a true vacuum and one local minimum $(\sigma_F,\phi_F)$
corresponding to a false vacuum.
We consider the situation in which the universe is initially
trapped at the false vacuum, and tunnels to the true vacuum
through the potential barrier.
This tunneling produces a true vacuum bubble
in which slow-roll inflation takes place.

The Lorentzian action $S[\sigma,\phi,g_{\mu\nu}]$ for our system is
  \begin{align}
   S[\sigma,\phi,g_{\mu\nu}]&=
   \int d^4x\sqrt{-g}
\left(\frac{m_{pl}^2}{2}R
-\frac{1}{2}g^{\mu\nu}\partial_\mu\sigma\partial_\nu\sigma
   -\frac{1}{2}g^{\mu\nu}\partial_\mu\phi\partial_\nu\phi
   -V(\sigma,\phi)\right)\,,
   \label{eq:multi_116}
  \end{align}
where $R$ is the Ricci scalar.
As in the CDL instanton method, we look for an instanton solution,
that is, a non-trivial solution of the Euclideanized system with
appropriate boundary conditions (see below).
The Euclidean action is obtained
by Wick-rotating the time coordinate, $S_E=-iS[x^0\to ix_E^0]$.

Let $\{\,\bar{\sigma}(x)\,,\ \bar{\phi}(x)\,,\ \bar{g}_{\mu\nu}(x)\,\}$
be an instanton solution.
Then the tunneling probability per unit time per unit volume is given by
  \begin{align}
   \Gamma =&A \exp\left(-B\right)\,,\nnmb
   B\equiv & S_E[\bar{\sigma},\bar{\phi},\bar{g}_{\mu\nu}]
   -S_E[\bar{\sigma}_F,\bar{\phi}_F,\bar{g}_{\mu\nu,F}]\,,
   \label{eq:28}
  \end{align}
where $\{\bar{\sigma}_F,\bar{\phi}_F,\bar{g}_{\mu\nu,F}\}$
is the solution staying at the false vacuum,
$\bigl(\bar{\sigma}_F(x),\bar{\phi}_F(x)\bigr)=\bigl(\sigma_F,\phi_F\bigr)$
and $\bar{g}_{\mu\nu,F}(x)$ being the Euclidean de Sitter metric
with vacuum energy $V(\sigma_F,\phi_F)$, namely a 4-sphere of
radius $\sqrt{3/V(\sigma_F,\phi_F)}\,m_{pl}$.
The coefficient $A$ is typically given as $A=O(M^4)$ with 
$M$ being the characteristic energy scale of the system.
Thus an instanton which gives the smallest possible action,
$S_E[\bar{\sigma},\bar{\phi},\bar{g}_{\mu\nu}]$ or $B$,
dominates the tunneling.

In the case of the false vacuum decay without gravity,
it is shown for a wide class of potentials that an $O(4)$-symmetric 
instanton gives the smallest Euclidean action~\cite{Coleman:1977th}.
Although there is no such theorem when gravity is included,
by continuation from the weak gravity limit,
it seems reasonable to assume that the same is true even in the
case with gravity. Therefore we look for an $O(4)$-symmetric instanton, 
as in the CDL analysis~\cite{Coleman:1980aw}.

The metric of an $O(4)$-symmetric Euclidean spacetime takes the form,
\begin{align}
ds_E^2&
=g_{\mu\nu}dx^\mu dx^\nu =N^2(t)dt^2+a^2(t)d\Omega_{(3)}^2\,,
  \label{eq:multi_3}
\end{align}
where $d\Omega_{(3)}^2$ is the metric on the unit 3-sphere $S_3$.
An $O(4)$-symmetric instanton depends only on the radial coordinate $t$
and has the form,
$\{\sigma,\phi,a\}=\{\bar{\sigma}(t),\bar{\phi}(t),\bar{a}(t)\}$.
With the $O(4)$-symmetric assumption, the Euclidean action for 
Eq.~\eqref{eq:multi_116} reduces to
\begin{align}
 S_E[\sigma,\phi,a]&=2\pi^2
   \int dt a^3\left[
  \frac{1}{N}\left(
      \frac{1}{2}{\sigma'}^2+\frac{1}{2}{\phi'}^2
     -\frac{3}{m_{pl}^2}\frac{{a'}^2}{a^2}\right)
     +N\left(V(\sigma,\phi)
     -\frac{3}{m_{pl}^2}\frac{1}{a^2}\right)
     \right]\,,
     \label{eq:16}
\end{align}
where the prime (${~}'$) denotes a derivative with respect to $t$.

The variation of the Euclidean action with respect to $N$ gives 
the Hamiltonian constraint,
      \begin{align}
       \left(\frac{{\bar{a}}'}{\bar{a}}\right)^2-\frac{1}{\bar{a}^2}&=
       \frac{1}{3m_{pl}^2}
     \left(\frac{1}{2}{\bar{\sigma}}'{}^2+\frac{1}{2}{\bar{\phi}}'{}^2
       -V(\bar{\sigma},\bar{\phi})\right).
       \label{eq:multi_6}
      \end{align}
Because $N$ is a gauge degree of freedom, we take $N=1$ from now on.
Then by varying the Euclidean action with respect to $\sigma$, $\phi$
and $a$, we obtain the equations of motion (EOM) for 
$\{\bar{\sigma}(t),\bar{\phi}(t),\bar{a}(t)\}$ as
\begin{align}
       \frac{{\bar{a}}''}{\bar{a}}+\frac{1}{3m_{pl}^2}
       \left({\bar{\sigma}}'{}^2
    +\bar{\phi}'{}^2+V(\bar{\sigma},\bar{\phi})\right)
       &=0\nnmb
       {\bar{\sigma}}''+3\frac{{\bar{a}}'}{\bar{a}}{\bar{\sigma}}'
       -V_{\bar{\sigma}}(\bar{\sigma},\bar{\phi})&=0\nnmb
       {\bar{\phi}}''+3\frac{{\bar{a}}'}{\bar{a}}{\bar{\phi}}'
       -V_{\bar{\phi}}(\bar{\sigma},\bar{\phi})&=0\,,
\label{180535_26Jul10}
\end{align}
where $V_{\bar{\phi}}=\partial_{\bar\phi}V$
and $V_{\bar{\sigma}}=\partial_{\bar\sigma}V$.

We adopt the boundary conditions for an instanton as in the 
CDL method~\cite{Coleman:1980aw}.
Here it may be worth noting that there is a subtlety
in the boundary conditions in the CDL method.
Without gravity an instanton is a non-trivial classical solution
which approaches a false vacuum at infinity.
On the other hand, with gravity a Euclidean spacetime becomes compact
and hence the instanton cannot arrive at the false vacuum.
This is a crucial problem from the view point of the physical
interpretation of the instanton~\cite{Bousso:1998vz,Rubakov:1999ir,Gen:1999gi}.
However, this problem is beyond the scope of the present paper.

The boundary conditions are determined by the regularity of an instanton.
 From Eq.~\eqref{eq:multi_6}, we find there will be two zeros of
$\bar{a}(t)$. We choose them to be at $t=0$ and $t=t_{\mathrm{end}}$,
that is $\bar{a}(0)=\bar{a}(t_{\mathrm{end}})=0$.
At $t=0$ and $t=t_{\mathrm{end}}$, $\dot{\bar{a}}(t)$ should 
satisfy $\bar{a}'(0)=-\bar{a}'(t_{\mathrm{end}})=1$.
This asymptotic behavior of $\bar{a}(t)$ around $t=0$ and 
$t=t_{\mathrm{end}}$ guarantees the regularity of the metric there.
Then the regularity of the scalar fields is guaranteed by imposing 
 $\bar{\sigma}'(t)=\bar{\phi}'(t)=0$
at $t=0$ and $t=t_{\mathrm{end}}$.
To summarize, the boundary conditions are
 \begin{eqnarray}
&& \bar{a}(0)=\bar{a}(t_{\mathrm{end}})=0\,,
\quad
 \bar{a}'(0)=-\bar{a}'(t_{\mathrm{end}})=1\,,
\cr
&&
 \bar{\sigma}'(0)=\bar{\sigma}'(t_{\mathrm{end}})=0\,,
\quad
 \bar{\phi}'(0)=\bar{\phi}'(t_{\mathrm{end}})=0\,.
\label{eq:multi_8}
 \end{eqnarray}

With the EOM \eqref{180535_26Jul10} and the boundary 
conditions \eqref{eq:multi_8}, we are ready to solve for an instanton.
It should be noted that even in the single-field case
the existence of an instanton depends on the
form of the potential~\cite{Jensen:1983ac,Linde:1995xm}.
In our case, the existence of a multi-field instanton may
be confirmed only after explicitly
constructing an instanton for a given, specific potential.

\begin{figure}[t]
         \hspace*{-0.5cm}
       \begin{minipage}{0.49\hsize}
       \begin{center}
        \includegraphics[width=7cm]{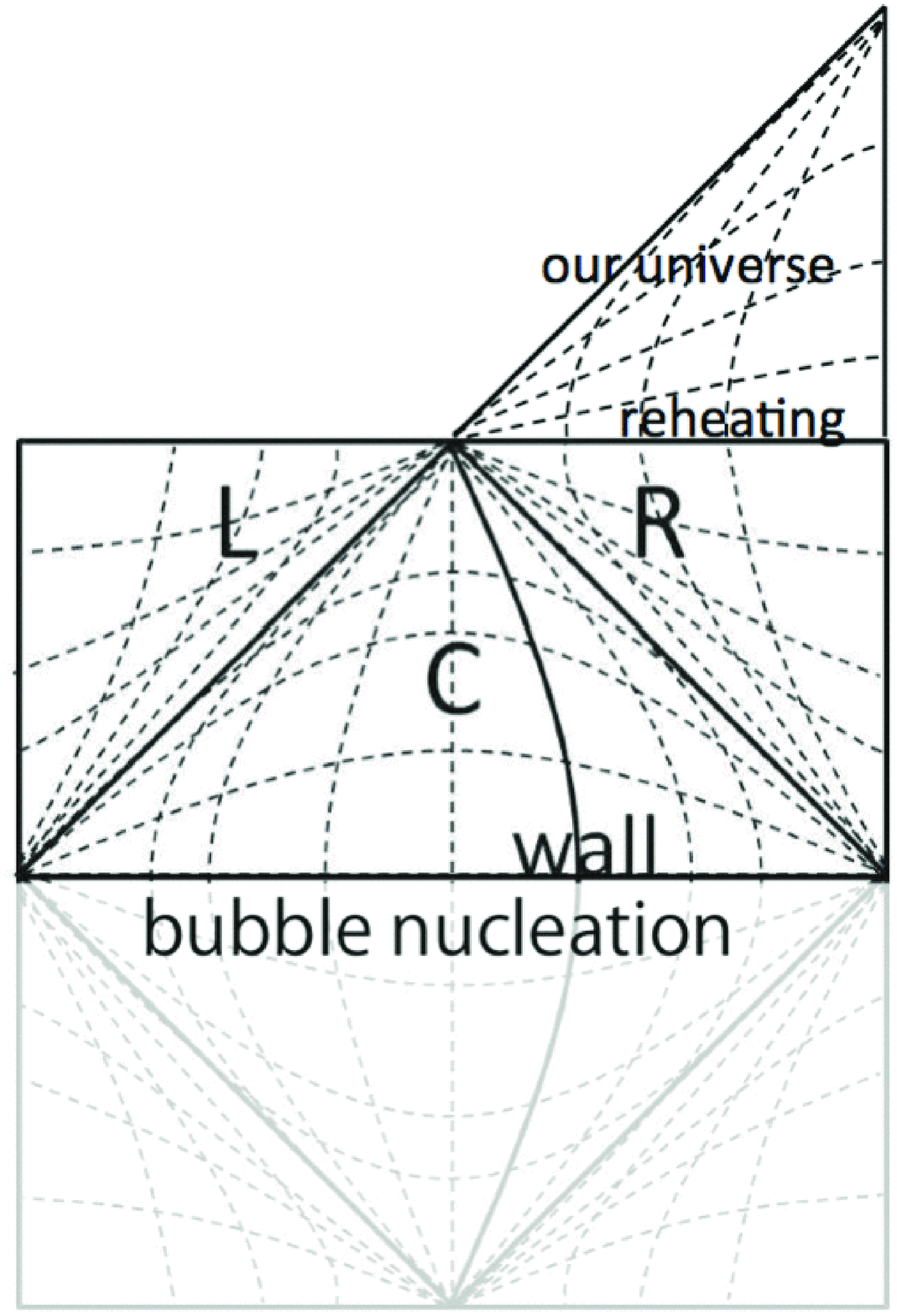}
 \caption{Penrose diagram of the universe after bubble nucleation.
Bubble nucleates on the Cauchy surface shown as a horizontal line. 
Region below the surface classically does not exist. 
The R-region is the open FLRW universe which 
experiences slow-roll inflation and reheating
to become our universe.}
    \label{fig:penrose_diag}
       \end{center}
   \end{minipage}
 \hspace*{+0.5cm}
       \begin{minipage}{0.49\hsize}
        \begin{center}
         \hspace*{-1cm}
 \includegraphics[width=9cm]{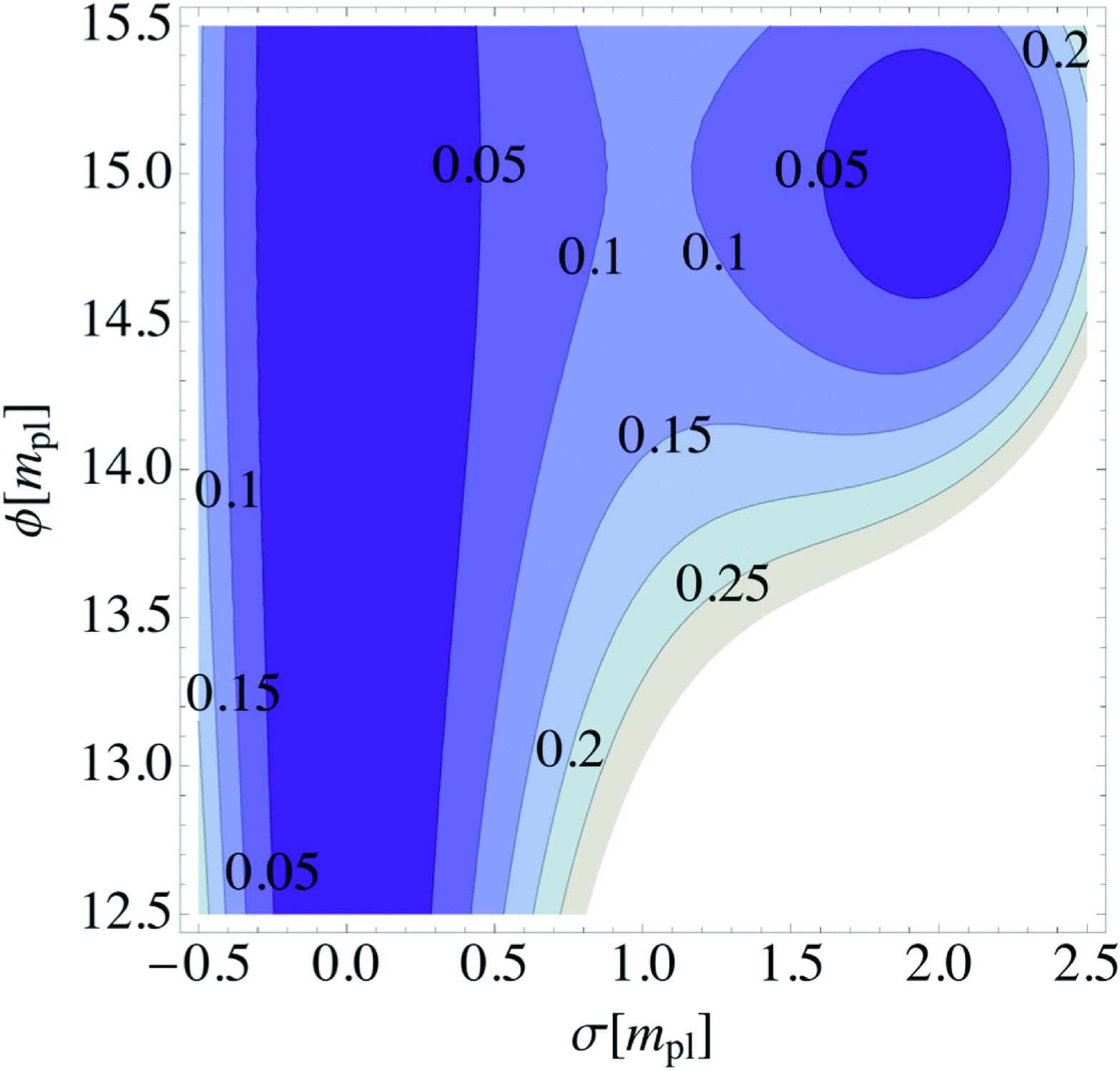}
\caption{Contour plot of 
$V(\sigma,\phi)=V_{\mathrm{tun}}(\sigma)+V_{\mathrm{infl}}(\sigma,\phi)$
for the values of the parameters,
$\alpha=0.1$, $\sigma_0=2m_{pl}$, $M_V=0.2m_{pl}$,
$\beta=0.1$, $m_\phi=10^{-6}m_{pl}$, $\phi_0=15m_{pl}$.
The contours are in Planck units.
The false vacuum is located near $(\sigma_0,\phi_0)= (2m_{pl},15m_{pl})$,
and the true vacuum is located at $(\sigma,\phi)=(0,0)$ which is not in
the graph. The scalar fields tunnel from the false vacuum to the 
other side of the potential barrier.
Then $\sigma$ undergoes rapid damped oscillations 
while $\phi$ slowly rolls down towards the true vacuum:
See Sec.~\ref{sec:find-inst-numer} and \ref{sec:slow-roll-after}.}
\label{fig:pot_contour}
        \end{center}
       \end{minipage}
\end{figure}

The spacetime geometry of the universe and 
the field configuration after tunneling are obtained
by analytical continuation of an instanton from the Euclidean
to Lorentzian regions~\cite{Coleman:1980aw}.
Here, we adopt the coordinate system used in \cite{Yamamoto:1996qq}
to describe the universe after tunneling.
The whole universe is divided to R, L and C regions 
as in Fig.~\ref{fig:penrose_diag}, which is a Penrose diagram of
the universe after bubble nucleation.
In the one-bubble open inflation scenario, our universe 
corresponds to the R region inside the bubble,
and we assume this part of the universe experiences reheating after 
inflation and becomes our hot universe.
The coordinates in the R, L and C regions are related with 
those in the Euclidean region by analytical continuation as
\begin{align}
 \begin{cases}
  t_R=-it&\left(0\leq t_R<\infty\right)\\
 r_R=i\chi&\left(0\leq r_R<\infty\right),\hspace{0.2cm} 
 \end{cases}
\begin{cases}
 t_L=i\left(t-t_{\mathrm{end}}\right)&\left(0\leq t_L<\infty\right)\\
 r_L=i\chi&\left(0\leq r_L<\infty\right),\hspace{0.2cm} 
\end{cases}
\begin{cases}
 t_C=t&\left(0\leq t_C\leq t_{\mathrm{end}}\right)\\
 r_C=i\left(\chi-\frac{\pi}{2}\right)&\left(0\leq r_C<\infty\right).
\end{cases}
\end{align}
The metrics in these regions are written as
\begin{align}
  ds_R^2&=-dt_R^2+a_R^2(t_R)\left(dr_R^2+\sinh^2r_R d\Omega^2\right)\,,
\nnmb
  ds_L^2&=-dt_L^2+a_L^2(t_L)\left(dr_L^2+\sinh^2r_L d\Omega^2\right)\,,
\nnmb
  ds_C^2&=dt_C^2+a_C^2(t_C)\left(-dr_C^2+\cosh^2r_C d\Omega^2\right)\,,
\label{eq:3}
\end{align}
where the scale factor in each region is given respectively by
$a_R(t_R)=i\bar{a}(-it_R)$,
 $a_L(t_L)=-i\bar{a}(-it_L+t_{\mathrm{end}})$
and $a_C(t_C)=\bar{a}(t)$.

We are interested in the region R which corresponds to our FLRW universe.
The scalar fields in the region R are also given by analytical continuation,
\begin{align}
 \sigma_R(t_R)=\bar{\sigma}(-it_R),\hspace{0.5cm}
 \phi_R(t_R)=\bar{\phi}(-it_R)\,.
\end{align}
The Lorentzian EOM for the scale factor and the scalar fields are
      \begin{align}
       \frac{\ddot{a}_R}{a_R}+\frac{1}{3m_{pl}^2}
       \left(\dot{\sigma}_R^2+\dot{\phi}_R^2-V(\sigma_R,\phi_R)\right)
       &=0\,,\nnmb
       \ddot{\sigma}_R+3\frac{\dot{a}_R}{a_R}\dot{\sigma}_R
       +V_{\sigma_R}(\sigma_R,\phi_R)&=0\,,\nnmb
       \ddot{\phi}_R+3\frac{\dot{a}_R}{a_R}\dot{\phi}_R
       +V_{\phi_R}(\sigma_R,\phi_R)&=0\,,
       \label{eq:21}
      \end{align}
where the dot ($\dot{~}$) denotes a derivative with respect to $t_R$.
The initial conditions at $t_R=0$ are given as
\begin{align}
 a_R(0)=0,\hspace{0.2cm}
 \sigma_R(0)=\bar{\sigma}(0),\hspace{0.2cm}
 \phi_R(0)=\bar{\phi}(0).
\end{align}

%%%%%%%%%%%%%%%%%%%%%%%%%%%%%%%%%%%%%%%%%%%%%%%%%%%%%%%%%%%%%%%%%%%% 
%%%%%%%%%%%%%%%%%%%%%%%%%%%%%%%%%%%%%%%%%%%%%%%%%%%%%%%%%%%%%%%%%%%% 
\section{Multi-field open inflation model}
\label{sec:model}
%%%%%%%%%%%%%%%%%%%%%%%%%%%%%%%%%%%%%%%%%%%%%%%%%%%%%%%%%%%%%%%%%%%% 
%%%%%%%%%%%%%%%%%%%%%%%%%%%%%%%%%%%%%%%%%%%%%%%%%%%%%%%%%%%%%%%%%%%% 
We now construct a concrete one-bubble multi-field open inflation model 
motivated by string landscape.
In Sec.~\ref{sec:potential}, we propose a two-field system
with a simple potential which satisfies requirements 
for a successful open inflation model.
Then, in Sec.~\ref{sec:find-inst-numer},
we explicitly obtain a multi-field instanton for the 
potential proposed in Sec.~\ref{sec:potential}.
Finally, in Sec.~\ref{sec:slow-roll-after}, using thus 
obtained instanton in Sec.~\ref{sec:find-inst-numer} as the initial condition,
we solve the evolution of the universe inside the bubble
until the end of slow-roll inflation.
We find that the one-bubble open inflation scenario is indeed
successfully realized in the system proposed in Sec.~\ref{sec:potential}.

%%%%%%%%%%%%%%%%%%%%%%%%%%%%%%%%%%%%%%%%%%%%%%%%%%%%%%%%%%%%%%%%%%%% 
\subsection{Potential}
\label{sec:potential}
%%%%%%%%%%%%%%%%%%%%%%%%%%%%%%%%%%%%%%%%%%%%%%%%%%%%%%%%%%%%%%%%%%%% 
Let us consider multi-field open inflation in a two-field system.
Inspired by \cite{Linde:1995xm}, we consider the case
where one of the scalar fields $\sigma$ plays a major role in
tunneling and and the other $\phi$ in slow-roll inflation inside the bubble.
To realize such a situation, we split the potential 
$V(\sigma,\phi)$ to two parts as
    \begin{align}
     V(\sigma,\phi)&=V_{\mathrm{tun}}(\sigma)+V_{\mathrm{infl}}(\sigma,\phi)\,,
     \label{eq:potential}
    \end{align}
and assume that $V_{\mathrm{tun}}(\sigma)$ essentially determines
the tunneling from the false vacuum,
and that $V_{\mathrm{infl}}(\sigma,\phi)$ dominates
the dynamics of slow-roll inflation after tunneling.
To stabilize the false vacuum, $V_{\mathrm{infl}}(\sigma_F,\phi)$ 
should give a large mass for $\phi$ around the false vacuum.
On the other hand, after the tunneling, $\sigma$ should reach its
true vacuum value $\sigma=\sigma_T$ sufficiently fast and 
$V_{\mathrm{infl}}(\sigma_T,\phi)$ should give a potential
for slow-roll inflation. 
We also assume that the number of $e$-folds of this slow-roll inflation 
is about $60$, as suggested by the string landscape
argument~\cite{Freivogel:2005vv}. This constrains the initial
value of $\phi$ after tunneling.

Now, we will construct a concrete form of the potential which 
satisfies the above requirements.
With regard to $V_{\mathrm{tun}}(\sigma)$,
we consider a simple quartic potential,
 \begin{align}
       V_{\mathrm{tun}}(\sigma)&=
       \alpha \sigma^2\left\{\left(\sigma-\sigma_0\right)^2+M_V^2\right\},
       \label{eq:tunpot}
 \end{align}
where the parameters are $\alpha$, $\sigma_0$ and $M_V$.
This potential is in the same form as the potential~\eqref{Vsigma},
but here we consider the range of the parameters in the context
of the string landscape.
The parameter $\alpha$ is a quartic self-coupling constant
which we assume $\alpha=O(1)$.
In the string landscape, 
the characteristic energy scale is expected to be the Planck scale.
Hence we assume $\sigma_0$ and $M_V$ are $O(m_{pl})$.
In order to realize two minima in the potential $V_{\mathrm{tun}}(\sigma)$,
we require the condition, $M_V\lesssim\sigma_0$.

It is known that a CDL instanton for
 $V_{\mathrm{tun}}(\sigma)$ does not always exist
even when there is a false vacuum.
Roughly speaking the existence of a CDL instanton depends on whether
the absolute value of the second derivative of the potential
 $|V_{\mathrm{tun},\sigma\sigma}(\sigma)|$ is larger than 
the squared Hubble parameter at the top of the barrier
$H^2(\sigma)=V_{\mathrm{tun}}(\sigma)/(3m_{pl}^2)$,
or more precisely, $|V_{\mathrm{tun},\sigma\sigma}(\sigma)|>4H^2(\sigma)$ at 
$\sigma=\sigma_{top}$~\cite{Jensen:1983ac,Linde:1995xm}.
This condition can be understood by recalling that
the wall thickness of an instanton is
$\Delta r\sim |V_{\mathrm{tun},\sigma\sigma}(\sigma_{top})|^{-1/2}$
while the size of the Euclidean 4-sphere is $r\sim\pi/H(\sigma_{top})$.
In order for an instanton to be fitted in the Euclidean sphere,
we need $\Delta r<r$, which is the condition for the
existence of a CDL instanton.
When this condition is violated, instead of a CDL instanton 
we obtain HM instanton, 
which is a trivial classical path staying on the 
top of the potential barrier~\cite{Hawking:1981fz}.
Since it is not easy to realize open inflation successfully
when the system has only a HM instanton,
as mentioned in Sec.~\ref{sec:introduction},
we concentrate on the case where a CDL instanton exists in what follows.
Although in a multi-field system the condition for the existence of a CDL 
instanton may be modified from the single-field case,
here we impose $\sigma_0\lesssim 3m_{pl}$
from the condition of the existence of a single CDL instanton, 
\begin{eqnarray}
\left|V_{\mathrm{tun},\sigma\sigma}(\sigma_{top})\right|
>4H^2(\sigma_{top})\,.
\label{eq:CDLcond}
\end{eqnarray}
We will come back to this issue later in Sec.~\ref{sec:probabilty}.

Let us consider the part of the potential that is supposed
to govern the slow-roll inflation,
 $V_{\mathrm{infl}}(\sigma,\phi)$.
We again restrict it to be at most quartic.
With this requirement, we set
 \begin{align}
 V_{\mathrm{infl}}(\sigma,\phi)&=
 \frac{1}{2}m_\phi^2 \phi^2 + \frac{\beta}{2}\sigma^2\left(\phi-\phi_0\right)^2,
       \label{eq:infpot}
 \end{align}
where the parameters are $\beta$, $m_\phi$ and $\phi_0$.
We assume $\beta=O(1)$.

For the potential given by the sum of Eqs.~(\ref{eq:tunpot}) and
 (\ref{eq:infpot}), let us compute the positions of the true vacuum 
and the false vacuum. It is easy to see that the true vacuum is at
$(\sigma_T,\phi_T)=(0,0)$ where $V(\sigma_T,\phi_T)=0$.
On the other hand, since there is a coupling between $\sigma$ and $\phi$,
the exact position of the false vacuum is not so easy to obtain
analytically. 
Here we assume $M_V^2\ll \sigma_0^2$ and $m_\phi^2\ll \beta \sigma_0^2$.
Then one can employ a perturbative expansion,
and at leading order the false vacuum is found to be at
$(\sigma_F,\phi_F)\sim \left(\sigma_0,\phi_0\right)$
with the vacuum energy given by 
$V(\sigma_F,\phi_F)\sim \alpha \sigma_0^2 M_V^2+m_{\phi}^2\phi_0^2/2$
(see Appendix~\ref{sec:det_pot} for details).
For simplicity, we restrict ourselves to this case in the following.

In order not to produce too much curvature perturbations
from the slow-roll inflation, the inflaton mass, $m_{\phi}$, 
should be less than about $10^{-6}m_{pl}$~\cite{Komatsu:2010fb}.
For the number of $e$-folds of about $60$, 
the initial value of $\phi$ should then be about $15m_{pl}$.
Thus we take $\phi_0 \sim 15m_{pl}$ anticipating that the 
inflaton field does not move much during the tunneling.
This is justified by the calculation in Sec.~\ref{sec:find-inst-numer}.
The actual number of $e$-folds is also explicitly calculated in
Sec.~\ref{sec:slow-roll-after}.

As mentioned in the beginning of this section, we also require that 
the mass square of the inflaton at false vacuum
is larger than the Hubble square at the false vacuum, 
$V_{\phi\phi}(\sigma_F,\phi_F)\gg V(\sigma_F,\phi_F)/(3m_{pl}^2)$,
to ensure the picture of the false vacuum decay.
If the mass is smaller than the Hubble parameter at the false 
vacuum, the quantum fluctuations dominate the motion of
the scalar fields and the picture of the quantum tunneling 
from a false vacuum ceases to be valid. An investigation of
this case may be of interest but beyond the scope of the present paper.
This condition implies $\alpha M_V^2/m_{pl}^2 < \beta$
under the assumptions discussed above that $m_\phi<10^{-6}m_{pl}$,
 $\alpha=O(1)$, $\beta=O(1)$, $\phi_0=O(m_{pl})$ and $\sigma_0=O(m_{pl})$. 

In the construction of a potential for one-bubble open inflation,
two conditions are to be satisfied at the same time.
One is the condition for the tunneling through a CDL instanton,
and the other is the condition for slow-roll inflation after tunneling.
For a single-field system, a very artificial potential is necessary to
fulfill both conditions at the same time~\cite{Linde:1998iw}.
However, as discussed above and will be shown below,
both conditions can be relatively easily satisfied with 
a simple quartic potential in a two-field model,
by assigning the roles of tunneling and slow-rolling to two 
different fields. A contour plot of the potential 
$V(\sigma,\phi)=V_{\mathrm{tun}}(\sigma)+V_{\mathrm{infl}}(\sigma,\phi)$
for a set of the parameters that lead to a successful model of open 
inflation constructed in the following subsections
is depicted in Fig.~\ref{fig:pot_contour}.

%%%%%%%%%%%%%%%%%%%%%%%%%%%%%%%%%%%%%%%%%%%%%%%%%%%%%%%%%%%%%%%%%%%% 
\subsection{Calculation of an instanton}
\label{sec:find-inst-numer}
\begin{figure}[!htbp]
     \begin{center}
      \includegraphics[width=9cm]{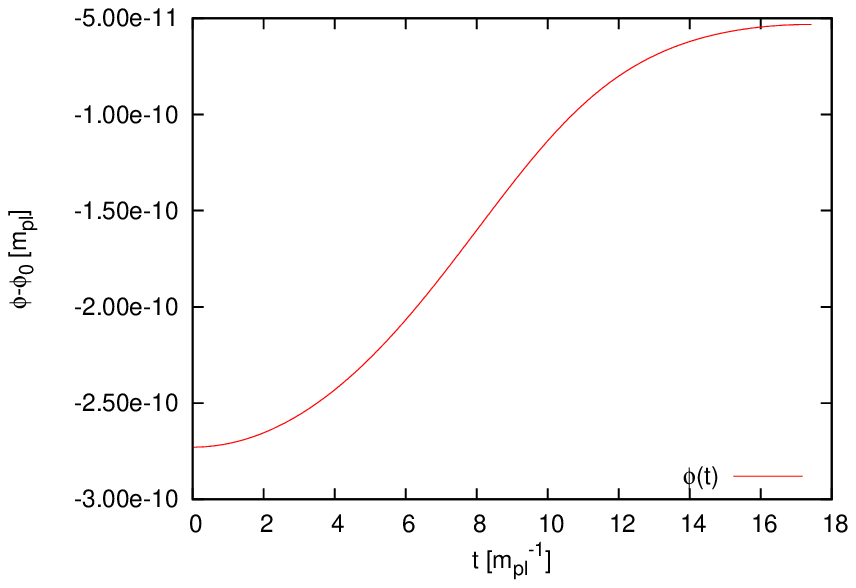}\\
      \hspace*{-2cm}      
      \includegraphics[width=9cm]{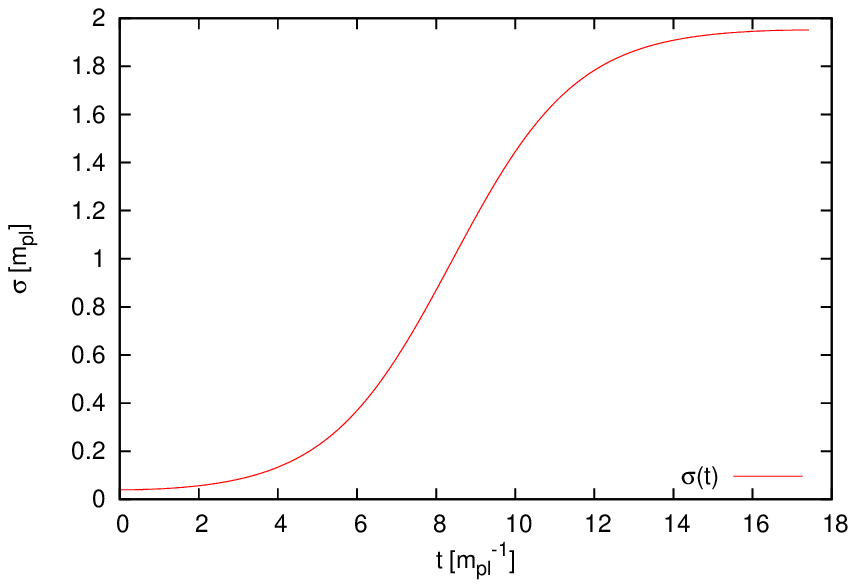}
      \includegraphics[width=9cm]{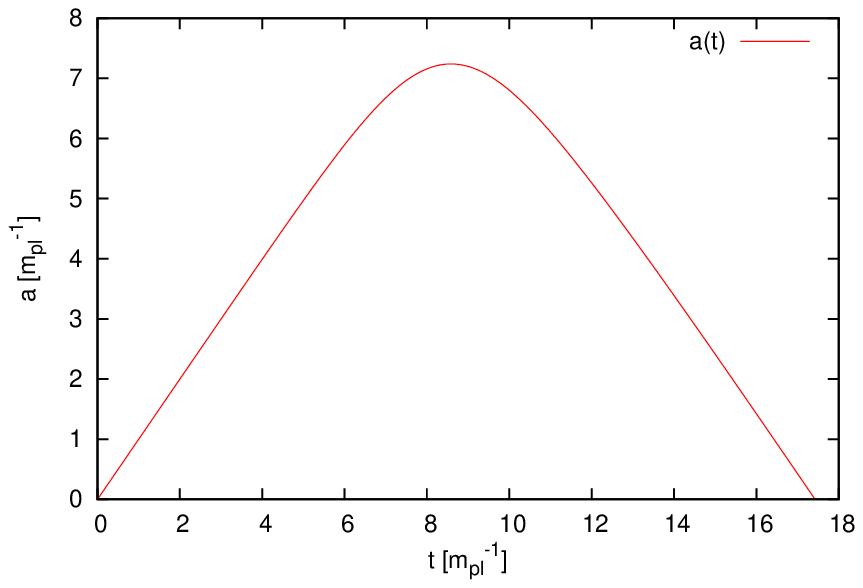}
      \caption{Multi-field instanton with gravity.
 The upper panel shows $\bar{\phi}(t)$. The lower-left and lower-right 
 panels show $\bar{\sigma}(t)$ and $\bar{a}(t)$, respectively.
 The potential is given by Eq.~\eqref{eq:potential}, \eqref{eq:tunpot}, 
 and  \eqref{eq:infpot}, with the parameters $\alpha=0.1$, $\sigma_0=2m_{pl}$,
 $M_V=0.2m_{pl}$, $ \beta=0.1$, $m_\phi=10^{-6}m_{pl}$, and $\phi_0=15m_{pl}$.}
      \label{fig:inst_all}
     \end{center}
    \end{figure}

%%%%%%%%%%%%%%%%%%%%%%%%%%%%%%%%%%%%%%%%%%%%%%%%%%%%%%%%%%%%%%%%%%%% 
We apply the extended CDL method formulated in Sec.~\ref{sec:formulation}
to our model proposed in Sec.~\ref{sec:potential}.
It should be noted that since the boundary conditions~\eqref{eq:multi_8}
are given at two different values of $t$, the existence of a non-trivial
 solution that satisfies the boundary conditions is confirmed only after
 one succeeds in explicitly constructing such a solution. 
We have searched for an instanton numerically with a shooting method
for the parameters $\alpha=0.1$, $\sigma_0=2m_{pl}$, $M_V=0.2m_{pl}$,
$\beta=0.1$, $m_\phi=10^{-6}m_{pl}$, $\phi_0=15m_{pl}$
and found an instanton.
The above parameter set satisfies the conditions discussed
in Sec.~\ref{sec:potential}.
The behavior of $\sigma$, $\phi$ and $a$ of thus
obtained instanton is shown in Fig.~\ref{fig:inst_all}.

From Fig.~\ref{fig:inst_all}, one sees that the scalar fields tunnel
to the state $(\bar{\sigma}(0),\bar{\phi}(0))\sim(0,\phi_0)$,
and the inflaton does not move significantly during tunneling
as expected in Sec.~\ref{sec:potential}.
Thus the inflaton starts rolling down from about $\phi_0=15m_{pl}$ 
inside the nucleated bubble.
The evolution of the universe after tunneling is explicitly 
calculated in Sec.~\ref{sec:slow-roll-after}.

%%%%%%%%%%%%%%%%%%%%%%%%%%%%%%%%%%%%%%%%%%%%%%%%%%%%%%%%%%%%%%%%%%%% 
\subsection{Evolution after the tunneling}
\label{sec:slow-roll-after}
%%%%%%%%%%%%%%%%%%%%%%%%%%%%%%%%%%%%%%%%%%%%%%%%%%%%%%%%%%%%%%%%%%%%
As discussed in Sec.~\ref{sec:formulation},
the universe inside the bubble is an open FLRW universe,
corresponding to the R-region in Fig.~\ref{fig:penrose_diag},
with the metric given by the second line of Eq.~\eqref{eq:3}.
We have numerically calculated the evolution inside the bubble
by solving the field equations~\eqref{eq:21}
with the initial conditions 
$a_R(0)=0$ and $(\sigma_R(0),\phi_R(0))=(\bar\sigma(0),\bar\phi(0))$
at $t_R=0$.

\begin{figure}[!tbp]
\begin{center}
\hspace*{-2cm}      
\includegraphics[width=9cm]{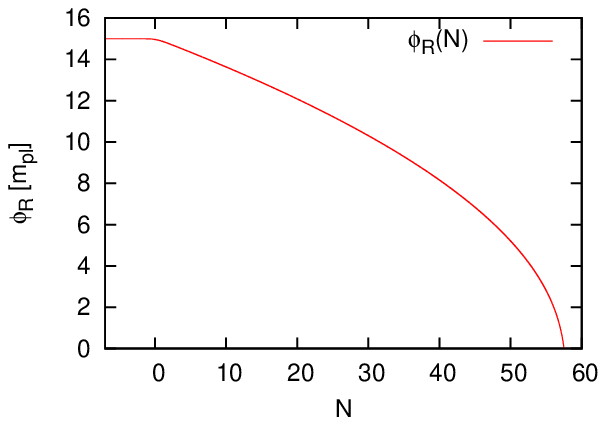}
\includegraphics[width=9cm]{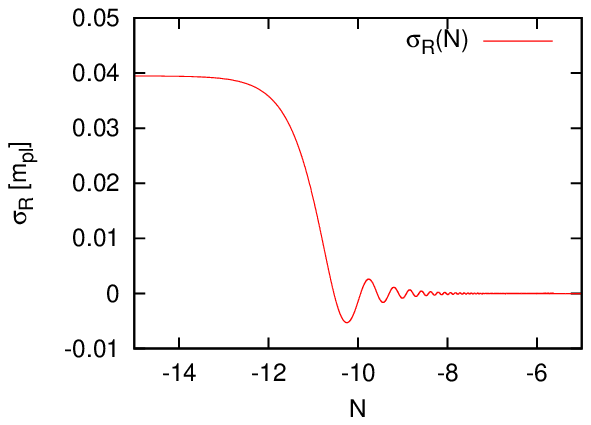}
\includegraphics[width=12cm]{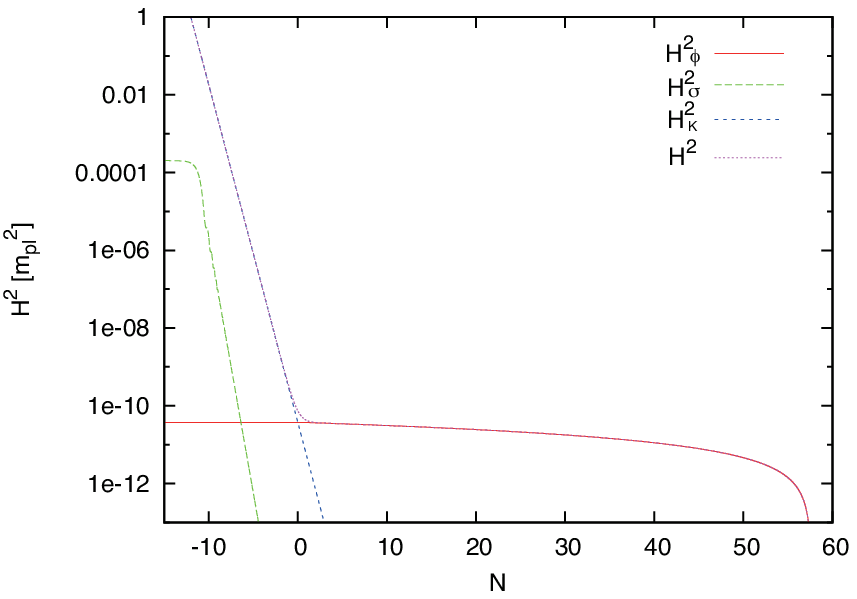}
\caption{Upper panels: The time evolution of the inflaton
 field $\phi$ (upper-left) and the tunneling field $\sigma$ (upper-right)
after tunneling as functions of $N\equiv\ln (a_R(t_R)/a_{R0})$
 with $a_{R0}$ being the value of the scale factor at the beginning of
 inflation.\\
Bottom panel: The contribution of the tunneling field $\sigma$,
 the inflaton field $\phi$, and the spatial curvature to the Hubble parameter
 as functions of $N$.
 $H_\phi$\,, $H_\sigma$ and $H_k$ are defined in Eq.~\eqref{eq:24}.
      }\label{fig:roll}
\end{center}
\end{figure}

The result is plotted in Fig.~\ref{fig:roll}.
In order to understand the evolution of the open universe,
we decompose the Hubble square $H^2\equiv (\dot{a}_R/a_R)^2$ into 
three contributions from $\phi$, $\sigma$ and the spatial curvature.
We denote these contributions by $H_\phi^2$, $H_\sigma^2$ and $H_k^2$, 
respectively. Their explicit definitions are
\begin{align}
 H_\phi^2&=\frac{1}{3m_{pl}^2}
\left(\frac{1}{2}{\dot\phi_R}^2+V\left(\sigma_T,\phi_R\right)\right)\,,\nnmb
 H_\sigma^2&=\frac{1}{3m_{pl}^2}\left(\frac{1}{2}{\dot\sigma_R}^2
+V\left(\sigma_R,\phi_R\right)-V\left(\sigma_T,\phi_R\right)\right)\,,\nnmb
 H_k^2&=\frac{1}{a_R^2}\,.
\label{eq:24}
\end{align}

Although the universe is curvature dominated right after tunneling,
the contribution from the spatial curvature, $H_k^2$, 
eventually decays as $a_R^{-2}$,
and the contribution from the inflaton field, $H_\phi^2$, starts to 
dominate, since
$H_\phi^2$ can be well approximated as a constant.
This happens at $a=a_{R0}$, where $a_{R0}\approx 1/H_\phi(0)$.
We define the number of $e$-folds as $N=\ln(a_R(t_R)/a_{R0})$.
With this choice of $a_{R0}$, the inflation starts at $N=0$.

One can see that the contribution from the tunneling field, 
$H_\sigma^2$, is larger than that from the inflaton field, 
$H_\phi^2$, and smaller than that from the spatial curvature, $H_k^2$,
right after tunneling. In the beginning $H^2$ decays as $1/a_R^2$,
and the mass of the tunneling field becomes larger than the Hubble 
parameter at $a_R\approx V_{{\rm tun},\sigma\sigma}^{-1/2}$.
After this epoch, $\sigma$ behaves as a non-relativistic matter 
and $H_\sigma^2$ decays rapidly as $a_R^{-3}$.
Thus, $H_\sigma^2$ becomes much smaller than $H_\phi^2$ 
before the curvature dominated era ends at $N=0$.
Thus $\sigma$ does not give any significant effect on the 
evolution of the universe.

Once $H_\phi^2$ becomes dominant, slow-roll inflation begins 
and lasts for about 60 $e$-folds.
We suppose that the standard reheating process occurs when 
inflation ends. 
Note that since the contribution of $\sigma$ to the dynamics
is small anyway, the evolution inside the bubble is similar
even if we choose the parameter set that gives
$H_\sigma^2 <H_\phi^2$ right after tunneling.

Before concluding this section, 
we emphasize that the solution given in Sec.~\ref{sec:find-inst-numer}
is the first explicitly obtained multi-field instanton with gravity 
for a simple potential, and
the evolution solved in Sec.~\ref{sec:slow-roll-after} gives
a successful model of open inflation where the multi-field dynamics 
is fully taken into account. This two-field model solves
the problem in single-field models in which one had to assume
quite an artificial potential~\cite{Linde:1998iw}.

In the past, two-field open-inflation models were studied
by several authors (see e.g. 
\cite{Linde:1995xm,GarciaBellido:1997te,Yamamoto:1996qq,Sasaki:1996qn}).
In particular, the quantum fluctuations of the scalar fields and
gravitational waves were discussed rather extensively.
However, the multi-field dynamics was not properly taken into account
in these previous studies.
Although they may perhaps be justified if regarded as 
results at leading order approximations,
because the roles of $\sigma$
and $\phi$ are fairly clearly separated,
quantitative verifications are left for future study.

%%%%%%%%%%%%%%%%%%%%%%%%%%%%%%%%%%%%%%%%%%%%%%%%%%%%%%%%%%%%%%% 
%%%%%%%%%%%%%%%%%%%%%%%%%%%%%%%%%%%%%%%%%%%%%%%%%%%%%%%%%%%%%%% 
\section{Effect of Multi-field Dynamics on Tunneling Rate}
\label{sec:probabilty}
%%%%%%%%%%%%%%%%%%%%%%%%%%%%%%%%%%%%%%%%%%%%%%%%%%%%%%%%%%%%%%% 
%%%%%%%%%%%%%%%%%%%%%%%%%%%%%%%%%%%%%%%%%%%%%%%%%%%%%%%%%%%%%%%
As one of the important effects of multi-field dynamics on
the tunneling, we consider its effect on the tunneling rate
$\Gamma\propto e^{-B}$. Related to this issue, we mention 
recent interesting papers on the tunneling rate in the context of
the cosmic string landscape.
Aguirre, Johnson and Larfors studied the effect of dilatonic coupling 
in multi-field systems~\cite{Johnson:2008vn,Aguirre:2009tp},
and concluded that it may completely prohibit tunneling in a
substantial part of the landscape.
Tye and Wohns studied the effect of resonant tunneling~\cite{Tye:2009rb},
and argued that it may drastically enhance the tunneling rate
and change the landscape. These effects surely deserve further study,
but here we focus on the system whose potential
form is defined in Sec.~\ref{sec:potential},
in which neither of the above effects is present.

To compare the tunneling rate in a single-field model with a multi-field
model is rather non-trivial, simply because there is no unique 
one-to-one correspondence between a single-field model and
a multi-field model. Here for a given two-field model, we compare
it with a single-field model obtained by fixing the inflaton
at the false vacuum $\phi=\phi_F$, 
\begin{eqnarray}
V^{(0)}(\sigma )\equiv V(\sigma ,\phi_F )\,,
\label{eq:singlelimit}
\end{eqnarray}
where $V(\sigma, \phi)$ is given by Eq.~(\ref{eq:potential}),
that is, the sum of $V_{\mathrm{tun}}(\sigma)$ 
and $V_{\mathrm{infl}}(\sigma,\phi)$ in Eqs.~\eqref{eq:tunpot}
and \eqref{eq:infpot}, respectively. 
We denote the instanton solution for this potential by
$\{\bar\sigma_0 (t),\bar\phi_0,\bar a_0(t)\}$,
where $\bar\phi_0=\phi_F$ and $\{\bar\sigma_0 (t),\bar a_0(t)\}$.
Then we consider the difference of the exponent of the tunneling rate
per unit volume per unit time between the two-field case 
and the single-field case,
\begin{align}
 \Delta B= B-B_0\,,
\end{align}
where $B=B[\bar\sigma ,\bar\phi ,\bar a]$ as defined in Eq.~\eqref{eq:28} 
and $B_0\equiv B[\bar\sigma_0 ,\bar\phi_0 ,\bar a_0]$.

To gain insight into the multi-field effect,
we then vary the inflaton mass $m_\phi$ in Eq.~(\ref{eq:infpot}).
This is because the inflaton would stay at the false vacuum
in the limit $m_\phi\to0$, while the inflaton dynamics during 
tunneling is non-negligible, if not substantial, for $m_\phi\neq0$.
As mentioned in Sec.~\ref{sec:potential}, for an one-bubble open inflation
model to be viable, it should not produce too large quantum fluctuations,
which implies the condition $m_\phi\lesssim 10^{-6}m_{pl}$.
However, here we tentatively ignore this constraint and 
consider the case $m_\phi\gg 10^{-6}m_{pl}$ as well,
in order to see the effect of the multi-field dynamics more clearly.

In Tables~\ref{tab:scl} and \ref{tab:scl2}, we list 
$\Delta \sigma=\bar{\sigma}(t_{\mathrm{end}})-\bar{\sigma}(0)$,
$\Delta \phi=\bar{\phi}(t_{\mathrm{end}})-\bar{\phi}(0)$, 
$B$, $B_0$, $B_{HM}$ and $\Delta B$,
for the parameters $\alpha=0.1$, $\sigma_0=2m_{pl}$, $M_V=0.2m_{pl}$
and $\phi_0=15m_{pl}$ with two different values of $\beta$;
$\beta=0.1$ and $0.01$. Here, $B_{HM}$ denotes the value of $B$
obtained for a HM instanton as a reference, where the action is 
evaluated at the saddle point between the false and true vacua,
$(\sigma_{top},\phi_{top})$, where 
$V_{\sigma}(\sigma_{top},\phi_{top})=V_{\phi}(\sigma_{top},\phi_{top})=0$.
It is listed there because as $m_\phi$ increases the CDL instanton 
approaches the HM instanton, and the CDL instanton ceases to exist
for sufficiently large $m_\phi$, as discussed below.

 From Tables~\ref{tab:scl} and \ref{tab:scl2},
we see that $\Delta B$ is either negative or too small to evaluate
within the accuracy of our computation.
This implies that the tunneling rate always increases 
when the effect of the multi-field dynamics is taken into account,
though the increase may be negligibly small in some cases.
The smallness of $|\Delta B|$ when $m_\phi\lesssim 10^{-2}m_{pl}$ 
is due to the smallness in the variation of $\bar{\phi}(t)$, or $\Delta\phi$,
during the tunneling process, as seen in the tables.

It should be noted that not only $\Delta B$ but also $B_0$ depends
on $m_\phi$ as well as on $\beta$, since 
the single-field case is given by the potential~(\ref{eq:singlelimit})
which depends both on $m_\phi$ and $\beta$.
As mentioned in the beginning of this section, this may be regarded
as a consequence of the non-uniqueness of a single-field model
for a given multi-field model. This fact makes it difficult to
make a quantitative statement about the effect of multi-field dynamics.
Nevertheless, the fact that the multi-field dynamics 
makes the tunneling rate larger is not affected.
Furthermore, if we fix the single-field case to be the limit
$m_\phi\to0$, which may be well approximated by the case of
$m_\phi=10^{-6}m_{pl}$, and compare it with cases of larger $m_{\phi}$,
we see that the decrease in $B$ is roughly inversely
proportional to $\Delta\phi$ when its effect is appreciable.

We also see from Tables~\ref{tab:scl} and \ref{tab:scl2}
that the CDL instanton approaches its corresponding HM instanton
as $m_\phi$ increases. This is because $V(\sigma_{top},\phi_{top})$ 
increases as $m_\phi$ increases, while the second derivative of 
$V(\sigma_{top},\phi_{top})$ along the tunneling path does not
change much. We know that the criterion for the existence of 
a CDL instanton in the single-field
case is given by Eq.~(\ref{eq:CDLcond}).
A straightforward extension of this criterion to 
the multi-field case is  
\begin{eqnarray}
|V''|>4H^2\,,
\label{eq:multicond}
\end{eqnarray}
where $V''$ is the second derivative along the steepest 
descent path at the saddle point,
or the negative eigenvalue of the mass matrix,
%\begin{eqnarray}
%\left.\left(\begin{array}{cc}
%V_{\sigma\sigma}&V_{\sigma\phi}\\
%V_{\phi\sigma}&V_{\phi\phi}
%\end{array}\right)\right|_{(\sigma_{top},\phi_{top})}\,,
%\end{eqnarray}
%that is,
\begin{eqnarray}
V''=\frac{V_{\sigma\sigma}+V_{\phi\phi}
-\sqrt{(V_{\sigma\sigma}-V_{\phi\phi})^2+4V_{\sigma\phi}^2}}{2}\,,
\end{eqnarray}
at the point $(\sigma_{top},\phi_{top})$, and 
\begin{eqnarray}
H^2=\frac{V(\sigma_{top},\phi_{top})}{3m^2_{pl}}\,.
\end{eqnarray}
The criterion (\ref{eq:multicond}) is found to be violated
at $m_\phi>4.12\times10^{-2}m_{pl}$ ($m_\phi>4.03\times10^{-2}m_{pl}$)
for $\beta=0.1$ ($\beta=0.01$).
In fact, we see that an instanton for $m_\phi=4\times10^{-2}m_{pl}$ is
very close to a HM instanton both from the smallness of $\Delta\sigma$
and the closeness of the value of $B$ to $B_{HM}$.
If we further increase $m_\phi$ beyond this critical value,
the potential barrier between the false and true vacua disappears
and the false vacuum ceases to exist.
The disappearance of the false vacuum is found to occur at 
$m_\phi=13.4\times10^{-2}m_{pl}$ ($m_\phi=8.1\times10^{-2}m_{pl}$)
for $\beta=0.1$ ($\beta=0.01$).

Qualitatively the fact that the multi-field dynamics 
increases the tunneling rate may be understood as follows.
As the variation of $\phi$ is small compared to that of $\sigma$
in our model, the single-field approximation is valid at leading order.
Now for a given single-field path 
$\{\sigma,a\}=\{\bar\sigma_0(t),\bar a_0(t)\}$,
the dynamics of $\phi$ is determined by minimizing the action 
$S_E[\bar\sigma_0(t), \bar a_0(t), \phi]$. Then the action for
a fixed value of $\phi$ would naturally be larger than that
minimizes the action $S_E[\bar\sigma_0(t), \bar a_0(t), \phi]$.
In the limit when the backreaction of the dynamics of $\phi$ on 
that of $\sigma$ and $a$ is small, the resulting value of the
action will be equal to the action of the actual CDL instanton.
Hence the effect of the multi-field dynamics is to decrease the
value of the action, and hence increase the tunneling rate.

Physically, by taking into account the dynamics of $\phi$,
the multi-field tunneling proceeds along the path where potential barrier 
is lower than the path along $\phi=\phi_F$.
Thus the tunneling occurs more easily when the multi-field dynamics
is taken into account. 
If we take the difference between the potential \eqref{eq:infpot}
and that at $\phi=\phi_F$, 
$\Delta V=V_{\mathrm{infl}}(\sigma,\phi)-V_{\mathrm{infl}}(\sigma,\phi_F)$,
we find that the potential energy along a multi-field tunneling path, 
which is away from $\phi_F$, decreases as $m_\phi$ increases
or $\beta$ decreases. This 
explains the dependence of $\Delta B$ on $m_\phi$ and $\beta$ 
in Tables~\ref{tab:scl} and \ref{tab:scl2}.

 \begin{table}
  \begin{tabular}{|c||c|c|c|c|c|c|}
\hline
 $m_\phi[m_{pl}]$ & $\Delta \sigma [m_{pl}]$ & $\Delta \phi [m_{pl}]$
 & $B$ & $B_0$ & $B_{HM}$ & $\Delta B$
\\ \hline
 $10^{-6}$ & $1.91$ & $2.20\times10^{-10}$
 & 12109.11 & 12109.11 & 12679.69 & $|\Delta B|<0.01$
\\ \hline
 $10^{-4}$ & $1.91$ & $2.20\times10^{-6}$
 & 12108.10 & 12108.10 & 12678.65 & $|\Delta B|<0.01$
\\ \hline
 $10^{-3}$ & $1.91$ & $2.19\times10^{-4}$
 & 12008.71 & 12008.71 & 12576.67 & $|\Delta B|<0.01$
\\ \hline
 $5\times10^{-3}$ & $1.90$ & $5.34\times10^{-3}$
 & 9975.05 & 9975.07 & 10484.43 & -0.02
\\ \hline
 $10^{-2}$ & $1.87$ & $1.97\times10^{-2}$
 & 6322.66 & 6322.85 & 6691.28 & -0.19
\\ \hline
 $2\times10^{-2}$ & $1.73$ & $6.00\times10^{-2}$
 & 2188.07 & 2189.41 & 2305.67 & -1.35
\\ \hline
 $3\times10^{-2}$ & $1.38$ & $8.73\times10^{-2}$
 & 849.20 & 852.07 & 868.55 & -2.87
\\ \hline
 $4\times10^{-2}$ & $0.49$ & $4.58\times10^{-2}$
 & 372.15 & 376.39 & 372.28 & -4.25
\\ \hline
 \end{tabular}
\caption{The values of several quantities of interest for
 different values of $m_\phi$.
 $\Delta \sigma=\bar{\sigma}(t_{\mathrm{end}})-\bar{\sigma}(0)$
 is the total variation of $\sigma$ during the tunneling,
 and $\Delta \phi=\bar{\phi}(t_{\mathrm{end}})-\bar{\phi}(0)$
 is that of $\phi$. $B$, $B_0$ and $B_{HM}$
 are the values of the bounce action for the full multi-field instanton,
 for the corresponding single-field instanton, and for the corresponding
 HM instanton, respectively, and $\Delta B=B-B_0$.
 The model parameters are
 $\alpha=0.1$, $\sigma_0=2m_{pl}$, $M_V=0.2m_{pl}$, $\phi_0=15m_{pl}$
 and $\beta=0.1$.
 }
  \label{tab:scl}
 \end{table}
 \begin{table}
 \begin{tabular}{|c||c|c|c|c|c|c|}
\hline
 $m_\phi[m_{pl}]$ & $\Delta \sigma [m_{pl}]$ & $\Delta \phi [m_{pl}]$
 & $B$ & $B_0$ & $B_{HM}$ & $\Delta B$
\\ \hline
 $10^{-6}$ & 1.91 & $3.20\times10^{-10}$ 
 & 12109.11 & 12109.11 & 12679.69 & $|\Delta B|<0.01$
\\ \hline
 $10^{-4}$ & 1.91 & $3.20\times10^{-6}$
 & 12108.10 & 12108.10 & 12678.65 & $|\Delta B|<0.01$
\\ \hline
 $10^{-3}$ & 1.91 & $3.20\times10^{-4}$
 & 12008.71 & 12008.71 & 12576.67 & $|\Delta B|<0.01$
\\ \hline
 $5\times10^{-3}$ & 1.90 & $7.72\times10^{-3}$
 & 9975.96 & 9976.06 & 10485.40 & -0.10
\\ \hline
 $10^{-2}$ & 1.87 & $2.79\times10^{-2}$
 & 6328.61 & 6329.83 & 6697.90 & -1.23
\\ \hline
 $2\times10^{-2}$ & 1.73 & $7.86\times10^{-2}$
 & 2196.44 & 2205.89 & 2316.09 & -9.44
\\ \hline
 $3\times10^{-2}$ & 1.37 & $1.04\times10^{-1}$
 & 844.13 & 865.98 & 863.91 & -21.84
\\ \hline
 $4\times10^{-2}$ & 0.24& $2.53\times10^{-2}$
 & 350.32 & 383.60 & 350.33 & -33.28
\\ \hline
 \end{tabular}
 \caption{Same as Table~\ref{tab:scl}, but for $\beta=0.01$.}
 \label{tab:scl2}
 \end{table}   

It is then natural to conjecture that the multi-field dynamics 
tends to increase the tunneling rate in general. The reason is
that although the action for a multi-field instanton is not
a minimum but a saddle point of the Euclidean action, there
is only a single negative direction that decreases the action.
In other words, if we can effectively separate out the dynamical 
degrees of freedom orthogonal to the tunneling path, the activation
of any of these degrees of freedom would increase the value of
the action. Hence freezing out these degrees of freedom by hand
would result in the increase of the action. Conversely the
inclusion of the dynamics of these degrees of freedom would
decrease the value of the action.

%%%%%%%%%%%%%%%%%%%%%%%%%%%%%%%%%%%%%%%%%%%%%%%%%%%%%%%%%%%%%%%%%%%% 
%%%%%%%%%%%%%%%%%%%%%%%%%%%%%%%%%%%%%%%%%%%%%%%%%%%%%%%%%%%%%%%%%%%% 
\section{Conclusion and Discussion}
 \label{sec:coclusion}
%%%%%%%%%%%%%%%%%%%%%%%%%%%%%%%%%%%%%%%%%%%%%%%%%%%%%%%%%%%%%%%%%%%% 
%%%%%%%%%%%%%%%%%%%%%%%%%%%%%%%%%%%%%%%%%%%%%%%%%%%%%%%%%%%%%%%%%%%% 
In this paper, motivated by string landscape
we have studied the dynamics of multi-field open inflation.
We have considered a system of two scalar fields,
in which one of the fields is to play a major role in quantum 
tunneling from a false vacuum, and is called the tunneling field,
and the other to play a major role in the subsequent slow-roll inflation
inside the bubble, and hence called the inflaton field.
For definiteness, we have considered the case when the
slow-roll inflation is of chaotic type, that is, a large-field model
of inflation. The introduction of two fields solves the difficulty 
in a single-field open inflation model in which
quite an artificial potential was necessary to
realize both tunneling and the subsequent slow-roll
inflation~\cite{Linde:1998iw}.
In our model the potential is a simple polynomial of quartic
order in the scalar fields. No fine-tuning of the parameters
has been necessary, although a certain degree of tuning has
been needed to satisfy the constraint that
the number of $e$-folds of inflation should be about 60,
as well as the standard requirement that the amplitude of
scalar perturbations should not exceed $10^{-5}$.

We have solved the Euclidean equations of motion numerically
to obtain the multi-field instanton explicitly.
Then by analytically continuing the instanton solution,
we have solved the evolution inside the bubble. 
We have confirmed that our model is a viable open inflation
model which can give the present curvature parameter 
$\Omega_{K,0}\sim10^{-2}-10^{-3}$.
Thus our model is the first concrete, viable model of open 
inflation with a simple potential, if not realistic.

Then in order to understand the effect of the multi-field dynamics
on quantum tunneling, we have considered the tunneling rate.
For a given two-field model, we have considered a corresponding
single-field model by fixing the value of the inflaton at the
false vacuum value, and compared the tunneling rates of
the two models. We have found that the multi-field dynamics
always increases the tunneling rate provided that 
the multi-field effect is perturbative.
This may be understood physically as a result of
the fact that the multi-field tunneling occurs along a classical path 
where the potential barrier is lower than the corresponding
single-field case in which the value of one of the fields is
fixed by hand. 

It should be noted that in \cite{Johnson:2008vn,Aguirre:2009tp} 
it was concluded that multi-field tunneling between two vacua by
an O(4)-symmetric instanton can be totally prohibited.
However, as mentioned in Sec.~\ref{sec:introduction},
this apparently opposite result can be understood by noting
the difference in the model considered there from ours.
This can be regarded as an example of highly non-trivial aspects of 
the multi-field tunneling. In order to understand the string landscape 
we definitely need further studies of multi-field tunneling.

In this paper, we have straightforwardly extended the CDL instanton method 
for the single-field tunneling with gravity to the multi-field case.
However, the CDL instanton method itself has some subtle points
as mentioned in Sec.~\ref{sec:formulation}.
One of the most important issues is the physical interpretation
of a CDL instanton.
Since the topology of a CDL instanton is $S^4$
the boundary conditions are determined by the regularity
of the solution. Hence there is no region in the solution
where the false vacuum is attained.
This is in marked contrast with the flat space case in 
which the topology is $E^4$ and the instanton approaches 
the false vacuum asymptotically at infinity.
This raises a question about the physical interpretation of an CDL instanton
that if it really describes the false vacuum decay.
In particular, in the limiting case a CDL instanton ceases to exist
and there is only a HM instanton which sits at the top
of the barrier. In this case, the analytic continuation of the solution
does not give the classical evolution after tunneling unless quantum
fluctuations are taken into account. 

It is important to discuss the observational implications of
multi-field open inflation models.
As far as the model we have constructed is concerned, 
the dynamics of the tunneling field turns out to be unimportant
inside the bubble. That is, the dynamics inside the bubble is
essentially that of single-field chaotic inflation.
Then we can calculate the power spectrum of the quantum fluctuations
by using the known formulation in the literature~\cite{Garriga:1997wz} 
without any modification. The expected result is that there is a slight 
suppression of the curvature perturbation on scales comparable to the
 curvature scale.
However, since $\Omega_{K,0}\lesssim 10^{-2}$, this effect on CMB, say,
is probably buried under the cosmic variance.

For models in which the energy scale of the false vacuum is
much higher, closer to the Planck scale, it was discussed that
the trace of the false vacuum could be observationally 
detected~\cite{Yamauchi:2011qq}. It remains to be seen if such
a model can be constructed in the multi-field context with 
relatively a simple potential.

Another interesting possibility is a model in which the tunneling 
field would remain
non-trivial inside the bubble to produce isocurvature perturbations
or in which damped oscillations of the tunneling field is delayed
to induce interesting features in the power spectrum.

Furthermore, since the quantum state inside the bubble is
known to be modified from the Bunch-Davis vacuum in 
general~\cite{Yamamoto:1996qq,Garriga:1997wz},
it is interesting to see if there is any multi-field model
in which this modification can be observable, say, in the CMB 
bispectrum~\cite{Meerburg:2009ys}.

If any of these possible models can be constructed in the context
of string landscape, it will provide a good observational test of
the string landscape. We hope to come back to these issues in
the near future.

%%%%%%%%%%%%%%%%%%%%%%%%%%%%%%%%%%%%%%%%%%%%%%%%%%%%%%%%%%%%%%%%%%%%%%%%%%%%%%%
%%%%%%%%%%%%%%%%%%%%%%%%%%%%%%%%%%%%%%%%%%%%%%%%%%%%%%%%%%%%%%%%%%%%%%%%%%%%%%%

\begin{acknowledgments}
We thank T.~Tanaka for useful discussions and valuable comments.
This work was supported in part by Monbukagaku-sho 
Grant-in-Aid for the Global COE programs, 
``The Next Generation of Physics, Spun from Universality 
and Emergence'' at Kyoto University.
This work was also supported in part by JSPS Grant-in-Aid for
 Scientific Research (A) No.~21244033,
and by Grant-in-Aid for Creative Scientific Research No.~19GS0219.
KS and DY were supported by Grant-in-Aid for JSPS Fellows
 Nos.~23-3437, and 20-1117, respectively. 
\end{acknowledgments}

\appendix
%%%%%%%%%%%%%%%%%%%%%%%%%%%%%%%%%%%%%%%%%%%%%%%%%%%%%%%%%%%%%%%%%%%% 
%%%%%%%%%%%%%%%%%%%%%%%%%%%%%%%%%%%%%%%%%%%%%%%%%%%%%%%%%%%%%%%%%%%% 
 \section{properties of the potential}
 \label{sec:det_pot}
%%%%%%%%%%%%%%%%%%%%%%%%%%%%%%%%%%%%%%%%%%%%%%%%%%%%%%%%%%%%%%%%%%%% 
%%%%%%%%%%%%%%%%%%%%%%%%%%%%%%%%%%%%%%%%%%%%%%%%%%%%%%%%%%%%%%%%%%%%
In this Appendix, we briefly summarize the properties of the potential
defined in Eqs.~\eqref{eq:potential}, \eqref{eq:tunpot},
and \eqref{eq:infpot}. Let us recapitulate them,
\begin{eqnarray}
V(\sigma,\phi)&=&V_{\mathrm{tun}}(\sigma)+V_{\mathrm{infl}}(\sigma,\phi)\,;
\cr
\cr
&&V_{\mathrm{tun}}(\sigma)=
       \alpha \sigma^2\left\{\left(\sigma-\sigma_0\right)^2+M_V^2\right\}\,,
\cr
&&V_{\mathrm{infl}}(\sigma,\phi)=
 \frac{1}{2}m_\phi^2 \phi^2 + \frac{\beta}{2}\sigma^2\left(\phi-\phi_0\right)^2\,.
\label{eq:V0th}
\end{eqnarray}

 Since the potential is quartic and non-negative,
there is at least one minimum.
To see the conditions for the existence of two minima,
we calculate the first derivatives of the potential:
\begin{align}
&V_{\sigma}=2\sigma
	\biggl[
   \alpha\left\{2\left(\sigma -\sigma_0\right)^2
    +\sigma_0\left(\sigma -\sigma_0\right)+ M_V^2\right\}
    +\frac{1}{2}\beta\left(\phi -\phi_0\right)^2
	\biggr]
\,,
\cr
&  V_{\phi}=m_\phi^2\phi +\beta\sigma^2\left(\phi -\phi_0\right)
\,,
\label{eq:V1st}
\end{align}
and the second derivatives:
\begin{align}
&V_{\sigma\sigma}=12\alpha
	\biggl[\sigma\left(\sigma -\sigma_0\right)
 	 +\frac{1}{6}\left(\sigma_0^2 +M_V^2\right)
    \biggr]
     +\beta\left(\phi -\phi_0\right)^2
\,,\cr
&V_{\sigma\phi}=2\beta\sigma \left(\phi -\phi_0\right)
\,,\cr
	&V_{\phi\phi}=m_\phi^2 +\beta\sigma^2
	\,.
\label{eq:V2nd}
\end{align}
It is easy to see that there is a minimum 
at $(\sigma ,\phi )=(0,0)$.
Since $V\geq 0$, it is the global minimum.

On the other hand, it is not easy to spell out the condition for
the existence of a false vacuum analytically. In general
we have to resort to a numerical method.
Nevertheless, if the conditions $M_V^2\ll \sigma_0^2$ 
and $m_\phi^2\ll\beta\sigma_0^2$ are both satisfied,
we find by an inspection of Eqs.~(\ref{eq:V1st}), together with
Eqs.~(\ref{eq:V2nd}), that
there is a local minimum at 
\begin{eqnarray}
(\sigma,\phi)
=\left(\sigma_0\left[1+O\left(\frac{M_V^2}{\sigma_0^2}\right)\right],
 \phi_0\left[1+O\left(\frac{m_\phi^2}{\beta\sigma_0^2}\right)\right]
\right).
\end{eqnarray}
A more careful, perturbative analysis gives
the position of the false vacuum and the vacuum energy as
\begin{align}
\sigma_F&=\sigma_0\left[1-\frac{M_V^2}{\sigma_0^2}
-2\left(\frac{M_V^2}{\sigma_0^2}\right)^2
-\frac{\beta\phi_0^2}{2\alpha\sigma_0^2}
\left(\frac{m_\phi^2}{\beta\sigma_0^2}\right)^2
+O(\epsilon^3)\right]\,,
\nnmb
\phi_F&= \phi_0\left[1-\frac{m_\phi^2}{\beta\sigma_0^2}
+\left(\frac{m_\phi^2}{\beta\sigma_0^2}\right)^2
-2\left(\frac{m_\phi^2}{\beta\sigma_0^2}\right)
\left(\frac{M_V^2}{\sigma_0^2}\right)
+O(\epsilon^3)\right]\,,
\nnmb
V(\sigma_F,\phi_F)&=
\alpha\sigma_0^2M_V^2\left[1-2\frac{M_V^2}{\sigma_0^2}+O(\epsilon^2)\right]
+\frac{1}{2}m_\phi^2\phi_0^2
\left[1-2\frac{m_\phi^2}{\beta\sigma_0^2}+O(\epsilon^2)\right]\,.\,
\label{eq:fvapprox}
\end{align}
where we have assumed ${M_V^2}/{\sigma_0^2}=O(\epsilon)$
and ${m_\phi^2}/(\beta\sigma_0^2)=O(\epsilon)$ with $\epsilon$
being a small parameter.

The false vacuum disappears when the equations $V_\sigma=V_\phi=0$ 
cease to have a real solution other than $\sigma=\phi=0$.
Then from the first of Eqs.~(\ref{eq:V1st}), one finds
that, under the assumption $\epsilon\ll1$, a sufficient condition for
the existence of a false vacuum is 
\begin{eqnarray}
\left(\frac{\beta\phi_0^2}{\alpha\sigma_0^2}\right)
\left(\frac{m_\phi^2}{\beta\sigma_0^2}\right)\lesssim1
\quad
\leftrightarrow
\quad
m_\phi^2\phi_0^2\lesssim\alpha\sigma_0^4\,.
\end{eqnarray}
From the last expression of Eqs.~(\ref{eq:fvapprox}),
one sees that the contribution to the false vacuum energy 
from $V_{\mathrm{infl}}(\sigma,\phi)$ becomes comparable to 
that from $V_{\mathrm{tun}}(\sigma)$
when $\alpha\sigma_0^2M_V^2\sim m_\phi^2\phi_0^2$.
However, in the case of a model we constructed in Sec.~\ref{sec:model},
the parameters satisfy the inequality
 $\alpha\sigma_0^2M_V^2\gg m_\phi^2\phi_0^2$. Hence 
the inflaton potential $V_{\mathrm{infl}}$ plays only a minor
role in the instanton solution, that is, in the quantum tunneling.

\end{document}